\documentclass{iau} 
\usepackage{graphicx,natbib,epstopdf,amsmath}

\newcommand{\apj}{ApJ}           

\newcommand{\aap}{A\&A}

\title[A Universal Density Profile]
{A `Universal' Density Profile for the Outer Stellar Halos of Galaxies}

\author[R.-S. Remus, A. Burkert \& K. Dolag]{Rhea-Silvia Remus$^1$, Andreas Burkert$^{1,2}$, \and Klaus Dolag$^{1,3}$}
\affiliation{$^1$Universit\"ats-Sternwarte M\"unchen, LMU, Scheinerstr.\ 1, D-81679 M\"unchen, Germany\\
$^2$MPI for Extraterrestrial Physics, Giessenbachstrasse 1, D-85748 Garching, Germany\\
$^3$MPI for Astrophysics, Karl-Schwarzschild Strasse 1, D-85748 Garching, Germany\\
email: {\tt rhea@usm.lmu.de} \\
}

\pubyear{2016}
\volume{321}  
\jname{Formation and evolution of galaxy outskirts}
\editors{A. Gil de Paz, J. Knapen \& J. Lee, eds.}
\begin{document}

\maketitle

\begin{abstract}
\looseness=-1
The outer stellar halos of galaxies contain vital information about the formation history of galaxies, since the relaxation timescales in the outskirts are long enough to keep the memory, while the information about individual formation events in the central parts has long been lost due to mixing, star formation and relaxation. To unveil some of the information encoded in these faint outer halo regions, we study the stellar outskirts of galaxies selected from a fully hydrodynamical high resolution cosmological simulation, called Magneticum. We find that the density profiles of the outer stellar halos of galaxies over a broad mass range can be well described by an Einasto profile. For a fixed total mass range, the free parameters of the Einasto fits are closely correlated. Galaxies which had more (dry) merger events tend to have lesser curved outer stellar halos, however, we find no indication that the amount of curvature is correlated with galaxy morphology. The Einasto-like shape of the outer stellar halo densities can also explain the observed differences between the Milky Way and Andromeda outer stellar halos.
\keywords{galaxies: halos -- structure, Galaxy: structure}
\end{abstract}

\firstsection 
\section{Introduction}
\looseness=-1
In addition to the clearly visible content of a galaxy, every galaxy is surrounded by a diffuse global stellar halo, which usually is assumed to be spherical.
This outer halo consists of old and thus not very massive stars, since there is basically no in-situ star formation in the outskirts of a galaxy because the gas density is much too low to form stars.
Therefore, these old populations must (mostly) be accreted through merging events and stripped from the main or the accreted galaxies during the encounters, as indicated for example by observations from \citet{martinez-delgado:2010}, or by the wealth of substructures observed around the Milky Way or Andromeda.
Measuring density and kinematic radial profiles of the outer stellar halos of galaxies could therefore provide a multitude of information about the mass accretion history and morphological changes of a galaxy.
We use the Magneticum Pathfinder simulations (\textbf{www.magneticum.org}, Dolag \textit{et al.}, in prep.) to analyse the shape and steepness of the density slopes of the outer halos of galaxies from Milky Way-mass galaxies up to brightest cluster galaxies (BCGs), and understand the implications about the accretion histories of galaxies that might be drawn from that.

\section{The Magneticum Pathfinder Simulations}
The Magneticum Pathfinder simulations (Dolag \textit{et al.}, in prep.) are a set of hydrodynamical cosmological boxes of different volumes and resolutions.
\begin{table}
\begin{center}
\caption{Magneticum simulations: current status}
\label{tab:sims}
\begin{tabular}{l | cccccc}
     	& Box0 & Box1 & Box2b &	Box2 & Box3 & Box4 \\
\hline
[Mpc/$h$] & 2688 & 896  &	640   &	352  &	128 &	48 \\
mr 	&  $2\times4536^3$ & $2\times1526^3$ & -- & $2\times594^3$&$2\times216^3$& $2\times81^3$ \\ 	 
hr 	& --  & -- &  $2\times2880^3$& $2\times1584^3$&	$2\times576^3$&	$2\times216^3$ \\
uhr & --	& -- & -- & -- & $2\times1536^3 (z=2)$ & $2\times576^3$ \\ 
\end{tabular}
\end{center}
\end{table}
They include metal-dependent radiative cooling and star formation, kinetic winds from SNIa, SNII and AGB stars, and formation and evolution of black holes and their according feedback. 
Furthermore, several improvements for smoothed particle hydrodynamics were included to more accurately treat turbulence and viscosity \citep[see][for more details on the simulations]{hirschmann:2014,teklu:2015}.
This results in a self-consistent formation of AGN populations, intra-cluster/group medium and galaxy populations which successfully reproduce observed properties \citep{remus:2015,teklu:2015,remus:2016}.
For this study we use galaxies selected from Box2 hr ($m_\mathrm{gas}=1.4\times10^8M_\odot/h$) and Box4 uhr ($m_\mathrm{gas}=7.3\times10^6M_\odot/h$) to cover the whole range of masses from Milky Way mass galaxies to BCGs.
In all simulations we use $h=0.704$ and each gas particle can spawn up to four stellar particles.

\section{A Universal Profile for Outer Stellar Halos}
\begin{figure}[t!]
\centering
\includegraphics[width=\textwidth]{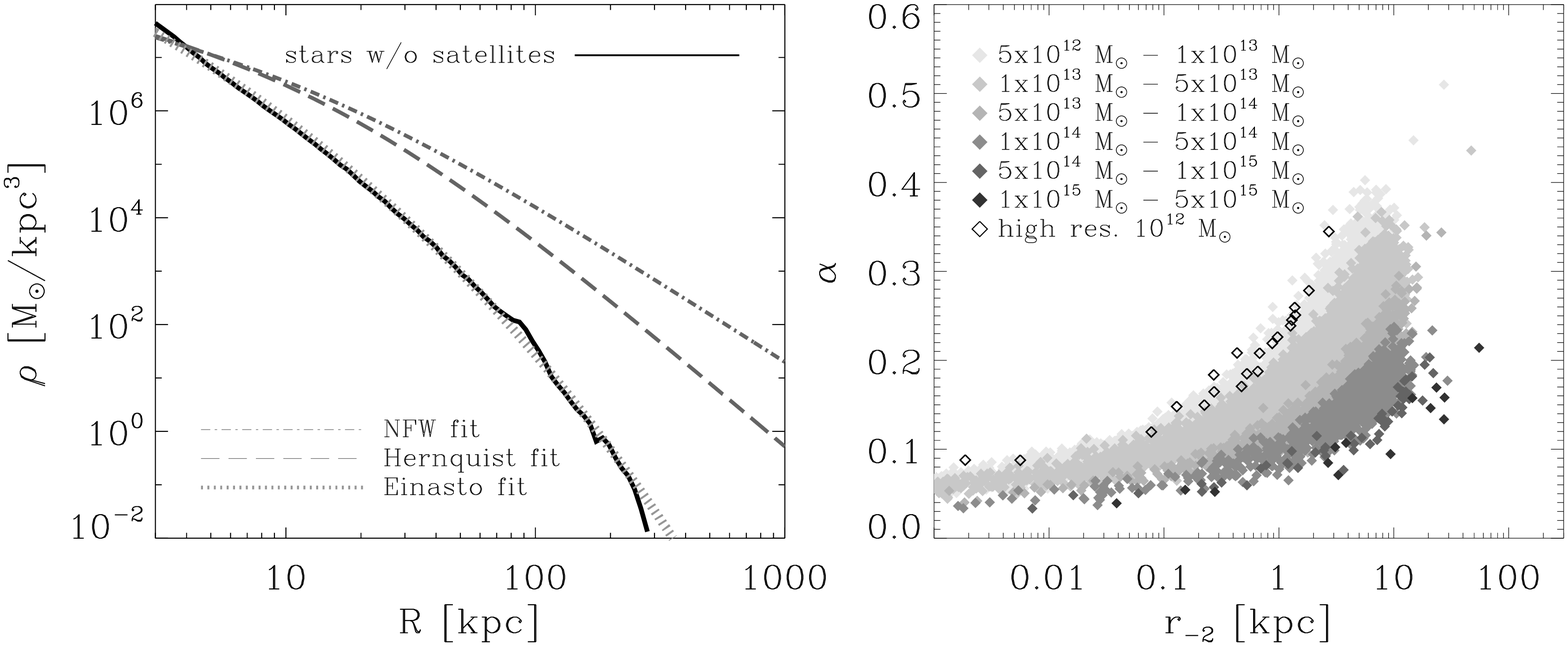}
\caption{\textit{Left:} Stacked stellar radial density profile (black solid line) for 24 Milky Way mass galaxies from Box4 uhr. NFW, Hernquist and Einasto fits to this profile according to legend.
\textit{Richt:} Einasto parameters $\alpha$ versus $r_\mathrm{-2}$ for the individual fits to stellar outer halos of galaxies from Box4 (MW mass galaxies, open diamonds) and Box2 (filled diamonds), covering different total masses (see legend).}
\label{fig:1}
\end{figure}

For stellar systems, there exists a number of density profiles in the literature that are used to analytically describe collisionless systems, most prominent the Hernquist profile \citep{hernquist:1990}.
For dark matter halos, the most commonly used profile is the NFW profile \citep{navarro:1996}.
The \textit{outer stellar halo density profiles} cannot successfully be described by any of these standard profiles, as shown in the left panel of Fig.~\ref{fig:1}, as the outer stellar halo is much stronger curved than any of the standard profiles.
In fact, a profile with an additional free parameter to represent the curvature of the outer stellar halo density profile is needed.
One such profile is the Einasto profile, first introduced by \citet{einasto:1965}:
\begin{flalign}
\rho(r) = \rho_{-2}~\exp \left\{ -\frac{2}{\alpha}~\left[~\left(\frac{r}{r_{-2}}\right)^{\alpha} -1\right]~\right\},
\end{flalign}
where $\alpha$ controls the curvature, $r_{-2}$ is the radius at which $\rho(r) \propto r^{-2}$, and $\rho_{-2}$ is the density within $r_{-2}$, see \citet{retana:2012}.
We fit Einasto profiles to the outer stellar halo density profiles for Milky Way-mass galaxies from Box4 uhr, and to group and cluster central galaxies from Box2 hr, as the former does not include enough massive central galaxies for this analysis due to the small box size of the simulation. 
We find the fitting parameters of these Einasto profiles to not be independent but rather closely correlated for fixed total mass over all tested mass ranges, as shown in the right panel of Fig.~\ref{fig:1}.
Interestingly, this correlation is independent of the galaxy type.
This self-similarity between the outer stellar halos over a broad range of masses strongly indicates that the formation of the outer stellar halo is dominated by accretion, which occurs in all halos.
The more dry accretion a galaxy underwent the less curved is the Einasto profile.

\section{Application to the Milky Way and Andromeda}
\looseness=-1
Observationally, there seems to be a clear difference between the outer stellar halos of the Milky Way and Andromeda: 
Fitting power laws $\rho(r) \propto r^\gamma$ to the outer stellar halos at radii of $40 \gtrsim r \gtrsim 100~\mathrm{kpc}$, \citet{deason:2014} and others report a power-law slope of $\gamma \approx -6$ for the Milky Way, while \citet{ibata:2014} and others find a slope of $\gamma \approx -3$ for Andromeda. 
As the Einasto profile is strongly curved at larger radii, it can locally always be represented by power laws with varying slopes.
We use Milky Way-mass halos ($M_\mathrm{tot}\approx 10^{12}M_\odot$) from Box4 uhr and fit power laws to the density profiles at the same radius range as the observations. 
We find slopes of $-6<\gamma<-2$, with the majority of the galaxies having slopes of $-6<\gamma<-5$, indicating that the Milky Way is actually a `normal' case while Andromeda must have had multiple dry (minor) mergers in the past, enhancing its satellite content as well as flattening the outer stellar halo density profile.
Thus, this self-consistently explains the observed differences between the outer stellar halos of the Milky Way and Andromeda, attributing it to their different mass accretion histories.

\section{Conclusions}
\looseness=-1
The density profiles of the outer stellar halos of galaxies from MW-mass up to BCGs can be well described by Einasto profiles.
The free parameters of the Einasto profile fits are closely correlated for fixed total masses over all tested mass ranges.
For a fixed total galaxy mass, a lesser curvature indicates a larger amount of dry merger events. 
Thus, the amount of curvature of the outer stellar halo density profile is a diagnostic for the merging history of galaxies. 
This can also explain the differences in the slopes of the power-law fits found for observations of the outer stellar halos of the Milky Way and Andromeda.


\begin{thebibliography}{}
\bibitem[Deason \etal\ (2014)]{deason:2014}
{Deason, A.~J., Belokurov, V., Koposov, S.~E., \& Rockosi, C.~M.} 2014,
\textit{\apj}, 787, 30

\bibitem[Einasto (1965)]{einasto:1965}
{Einasto, J.} 1965,
\textit{Trudy Astrofizicheskogo Instituta Alma-Ata}, 5, 87

\bibitem[Hernquist (1990)]{hernquist:1990}
{Hernquist, L.} 1990,
\textit{\apj}, 356, 359

\bibitem[Hirschmann \etal\ (2014)]{hirschmann:2014} 
{Hirschmann, M., Dolag, K., Saro, A., \etal\ } 2014, 
\textit{MNRAS}, 442, 2304

\bibitem[Ibata \etal\ (2014)]{ibata:2014}
{Ibata, R.~A., Lewis, G.~F., McConnachie, A.~W. \etal\ } 2014,
\textit{\apj}, 780, 128
    
\bibitem[Mart{\'{\i}}nez-Delgado \etal\ (2010)]{martinez-delgado:2010} 
{Mart{\'{\i}}nez-Delgado, D., Gabany, R.~J., Crawford, K., \etal\ } 2010, 
\textit{AJ}, 140, 962

\bibitem[Navarro \etal\ (1996)]{navarro:1996}
{Navarro, J.~F., Frenk, C.~S., \& White, S.~D.~M.} 1996,
\textit{\apj}, 462, 563

\bibitem[Remus \etal\ (2015)]{remus:2015} 
{Remus, R.-S., \etal\ } 2015, 
\textit{IAU Proceedings Vol. 309}, Galaxies in 3D across the Universe, 145

\bibitem[Remus \etal\ (2016)]{remus:2016} 
{Remus, R.-S., Dolag, K., Naab, T., \etal\ } 2016, 
\textit{ArXiv}, 1603.01619

\bibitem[Retana-Montenegro \etal\ (2012)]{retana:2012}
{Retana-Montenegro, E., van Hese, E., Gentile, G., \etal\ } 2012,
\textit{\aap}, 540, A70

\bibitem[Teklu \etal\ (2015)]{teklu:2015} 
{Teklu, A. F., Remus, R.-S., Dolag, K., \etal\ } 2015, 
\textit{ApJ}, 812, 29

\end{thebibliography}
\end{document}